\documentclass{webofc}

\usepackage[varg]{txfonts}   
\usepackage{hyperref}
\usepackage{url}
\usepackage{combelow}
\usepackage{wrapfig}
\newcommand{\xT}{{\mathbf{x}_\perp}}
\newcommand{\trento}{\textsc{trento}}
\newcommand{\vhlle}{\textsc{vHLLE}}
\hypersetup{colorlinks=true,citecolor=blue,urlcolor=blue,linkcolor=blue}
\begin{document}
\title{Quantifying the degree of hydrodynamic behaviour in heavy-ion collisions}
%
%

\renewcommand{\d}{\mathrm{d}}
\newcommand{\W}{\mathrm{W}}

\author{\firstname{Sören} \lastname{Schlichting}\inst{1}\and \firstname{Clemens} \lastname{Werthmann}\inst{1,2,3}\fnsep\thanks{Speaker, \email{clemens.werthmann@ugent.be}}
}
\date{\today}

\institute{ Fakultät für Physik, Universität Bielefeld, D-33615 Bielefeld, Germany
\and
           Incubator of Scientific Excellence-Centre for Simulations of Superdense Fluids, University of Wrocław, pl. Maxa Borna 9, 50-204 Wrocław, Poland
\and
            Department of Physics and Astronomy, Ghent University, 9000 Ghent, Belgium
          }

\abstract{Exploiting the first measurements of the same ion species in O+O collisons at RHIC and LHC, we propose an experimentally accessible observable to distinguish whether collective behaviour builds up through a hydrodynamic expansion of a strongly interacting QGP or through few rescatterings in a non-equilibrated dilute medium. Our procedure allows to disentangle the effects of the initial state geometry and the dynamical response mechanism on the total resulting anisotropic flow. We validate the ability of our proposed observable to discriminate between systems with different interaction rates using results from event-by-event simulations in kinetic theory in the Relaxation Time Approximation (RTA). As a proof of concept, we extract the degree of hydrodynamization for Pb+Pb collisions at LHC from experimental data.
}
\maketitle
\vspace{-15pt}
\section{Introduction}
\label{sec:intro}

The discovery of anisotropic flow in small collision systems has led to discussions over the degree to which these systems hydrodynamize. While anisotropic flow responses are strongest in equilibrated media, some flow can also be generated in dilute systems. It is not simply the presence but rather the magnitude of flow responses that indicates hydrodynamic behaviour. However, anisotropic flow is generated in response to anisotropies in the initial state geometry, which is poorly constrained. This prevents direct extraction of the magnitude of flow responses from experimental data.

We propose observables to disentangle information on the initial state geometry and degree of hydrodynamization~\cite{Ambrus:2024hks,Ambrus:2024eqa}. The hydrodynamization observable makes use of the opacity dependence of the flow response $\kappa=v_2/\epsilon_2$, or rather its power law behaviour. At large interaction rates, this response saturates to a constant, $\kappa(\hat{\gamma})=\kappa_{\rm ideal}$, while at small interaction rates, it is linear in the opacity, $\kappa(\hat{\gamma})=\kappa'_{\rm LO}\hat{\gamma}$.

\section{Setup}
\label{sec:setup}

To validate our proposed procedure, we perform 2+1D event-by-event simulations of O+O collisions at energies of $200\,$GeV (RHIC) and $7\,$TeV (LHC) in both kinetic theory and hydrodynamics. The initial states were generated in \trento~by randomly sampling from a list of pre-generated nucleon positions \cite{Lim:2018huo}, which take correlations like $\alpha$-clustering into account. 

Our kinetic theory code solves the Boltzmann equation in conformal RTA
\begin{align}
    p^\mu\partial_\mu f = C_{\rm RTA}[f]=-\frac{p^\mu u_\mu}{\tau_R}(f-f_{\rm eq})\;,\quad \tau_R=5\eta/s\cdot T^{-1}\,
\end{align}
where $f=\d N/\d^3x\d^3p$ is the statistical phase space distribution function and $\eta/s$ is the (constant) specific shear viscosity. The effective temperature $T$ is determined locally from the energy-momentum tensor via Landau matching. This simplified description is not suitable for an accurate comparison to experimental data, but it allows to study qualitative behaviour in hadronic collisions. In conformal RTA, the dynamics depend only on a single dimensonless parameter, the opacity $\hat{\gamma}$~\cite{Kurkela:2019kip,Ambrus:2021fej}. It encodes dependencies on the specific shear viscosity $\eta/s$, the initial transverse Radius $R$ and the initial energy per unit rapidity $\d E_\perp/\d \eta$.
\begin{align}
    \hat{\gamma}=\left(5\frac{\eta}{s}\right)^{-1}\left(\frac{1}{\pi a}R\frac{\d E_\perp}{\d \eta}\right)^{1/4}
\end{align}
We work with an equation of state $e=aT^4=\nu_{\rm eff}\frac{\pi^2}{30}T^4$, where we chose $\nu_{\rm eff}=42.25$ to be in agreement with lattice QCD results. In order to study the dynamics at various opacity scales with the same initial state geometry, we run simulations of all events at several values of $\eta/s$.

Simulations of second order Mueller-Israel-Steward type hydrodynamics are performed in the code \vhlle~\cite{Karpenko:2013wva} with a conformal equation of state and transport coefficients matched to RTA kinetic theory, in order to achieve a meaningful comparison to our kinetic theory results. At high interaction rates and close to equilibrium, hydrodynamics can be expected to agree with kinetic theory. However, large deviations persist in the pre-equilibrium period even at high opacities. In practice, this is typically circumvented by using a hybrid time evolution with a non-equilibrium description for early times. We instead choose to initialize hydrodynamics at early times and rescale the initial condition~\cite{Ambrus:2022qya,Ambrus:2022koq}. The rescaling scheme counteracts differences in pre-equilibrium by estimating them in local Bjorken flow.

We study elliptic flow in terms of the observable $\varepsilon_p$, which is defined via
\begin{align}
    \varepsilon_p\,e^{2i\Psi_p}=\frac{\int_\xT T^{xx}-T^{yy}+2iT^{xy}}{\int_\xT T^{xx}+T^{yy}}\;.
\end{align}
This observable measures flow of energy instead of particle number and has the advantage that it can be directly computed from the energy-momentum tensor without the need of particlization. It has been argued that $\varepsilon_p$ and the more commonly used harmonic flow coefficient $v_2$ can be related to each other via a simple $\sqrt{s_{\rm NN}}$-dependent conversion factor~\cite{Kurkela:2019kip}.

\section{Numerical Results}
\label{sec:results}

\begin{figure*}
    \centering
    \includegraphics[width=0.49\linewidth]{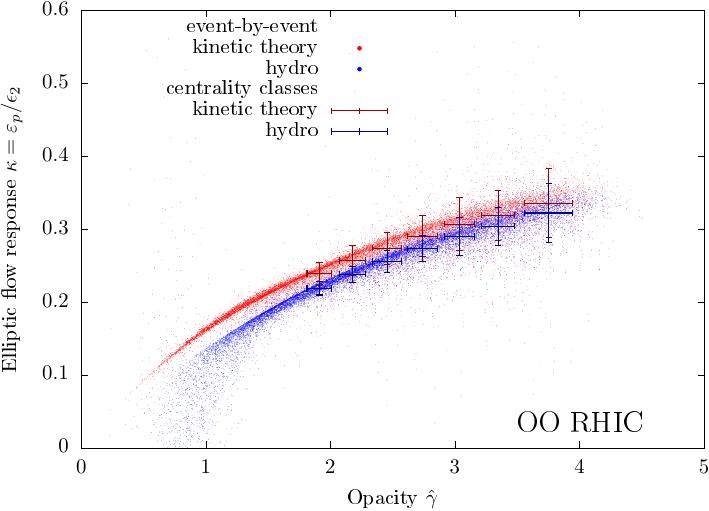}
    \includegraphics[width=0.49\linewidth]{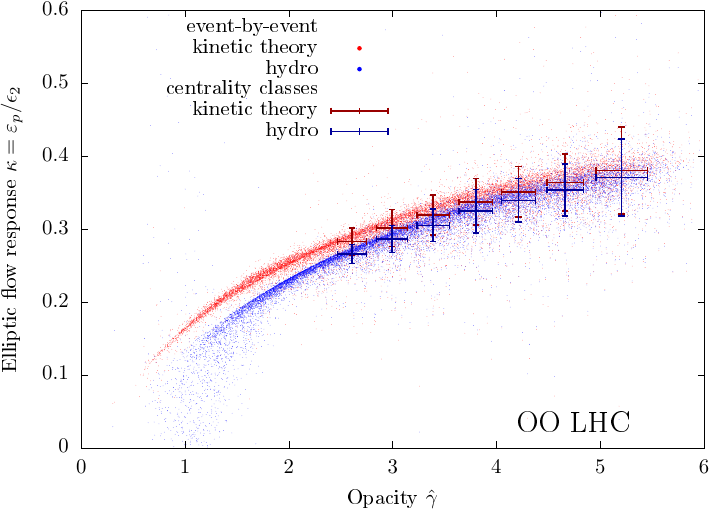}
    \caption{Event-by-event elliptic flow response $\kappa=\varepsilon_p/\epsilon_2$ as a function of the event-by-event opacity $\hat{\gamma}$. Hydrodynamic results in blue are compared to kinetic theory results in red. Crosses show mean and variance within centrality classes of size 10\%.}
    \label{fig:HYDROvsEKT}
    \vspace{-15pt}
\end{figure*}

The interpretation of our proposed hydrodynamization obervable relies on the previously established criterion for the applicability of hydrodynamics in terms of the opacity, $\hat{\gamma}\gtrsim 3$~\cite{Ambrus:2022qya,Ambrus:2022koq}. This criterion was extracted from comparisons of hybrid hydro+kinetic theory to full kinetic theory simulation results using an averaged initial state. To make sure that these insights transfer also to event-by-event simulations, we again compare results for the  elliptic flow response coefficient $\kappa=\varepsilon_p/\epsilon_2$ from hydrodynamic and kinetic theory based simulations in Fig.~\ref{fig:HYDROvsEKT}. While the geometry fluctuations do introduce some spread of $\kappa$ at fixed opacity $\hat{\gamma}$, it is clearly visible that in all cases the event-by-event values mostly follow common curves $\kappa_{\rm kin}(\hat{\gamma})$ and $\kappa_{\rm hyd}(\hat{\gamma})$. These curves have the same form in O+O collisions at RHIC and LHC, with the main difference being that LHC results reach to higher opacities. In fact, as can be seen in the left plot of Fig.~\ref{fig:W_sim}, the curves are universal to all collision systems. In agreement with the earlier results, the hydrodynamic flow response coefficient falls short of the one in kinetic theory at small opacities, but converges to it from below in the large opacity limit. We conclude that insights gained in the average event setup are still valid and that the degree of hydrodynamization of realistic collision systems can be gauged in terms of the opacity $\hat{\gamma}$.

\begin{figure*}
    \centering
    \includegraphics[width=0.49\linewidth]{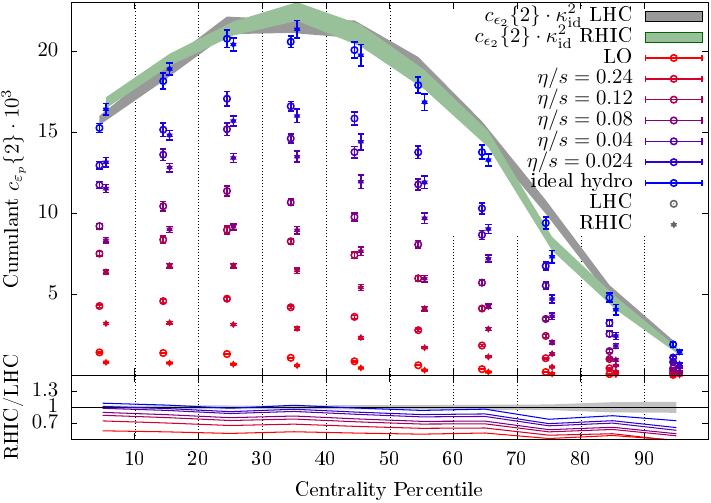}
    \includegraphics[width=0.49\linewidth]{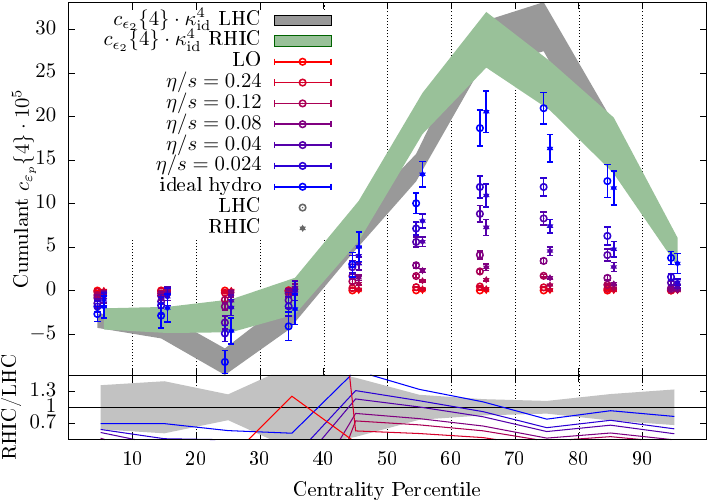}
    \caption{Second order (left) and fourth order (right) elliptic flow cumulants as a function of centrality, as computed from kinetic theory simulation data for O+O collisions at RHIC (stars) and LHC (open circles) at various values of the specific shear viscosity (different colors). Bands show the corresponsing cumulants for the initial state ellipticity.}
    \label{fig:RHICvsLHC}
    \vspace{-25pt}
\end{figure*}

Next, we compare the statistics of event-by-event elliptic flow at RHIC and LHC in terms of the second and fourth order flow cumulants $c_{\varepsilon_p}\{2\}=\langle\epsilon_p^2\rangle$ and $c_{\varepsilon_p}\{4\}=\langle\epsilon_p^4\rangle-2\langle\epsilon_p^2\rangle^2$.
Fig.~\ref{fig:RHICvsLHC} shows the centrality dependence of these cumulants as obtained from kinetic theory simulations for different values of the shear viscosity $\eta/s$, i.e. different opacity scales. The main effect of increasing the opacity (decreasing $\eta/s$) is an overall increase in the magnitude of elliptic flow responses, which translates to larger values of the cumulants. Going towards smaller shear viscosities, the flow response curve flattens as it asymptotes to the ideal hydrodynamic limit. In this limit, all events have roughly the same response coefficient and the form of the elliptic flow cumulant curve agrees with that of the corresponding cumulant for the initial state ellipticity $\epsilon_2$, plotted as green and grey bands. On the other hand, smaller or less energetic events have smaller opacities, which introduces a centrality dependence of the typical opacity. This means that in less central collisions the flow response coefficient will decrease faster with increasing $\eta/s$, introducing a modulation of the cumulant curve as a function of centrality. In particular, the maximum of the curve shifts towards more central classes. Similarly, RHIC events have smaller opacities than LHC events at the same shear viscosity and therefore have smaller flow responses, as can be seen in the bottom plots of the ratios. Again, this effect is more pronounced at larger shear viscosities, while in the ideal hydrodynamic limit this ratio goes to unity.

\section{Disentangling effects of geometry and response}
\label{sec:disentangling}

The similarity between centrality dependencies of the ideal hydrodynamic elliptic flow cumulants and initial state ellipticity cumulants can be expressed mathematically as follows:

\begin{align}
\langle(\varepsilon_p)^n\rangle=\langle(\kappa\epsilon_2)^n\rangle\approx \bar{\kappa}^n\langle(\epsilon_2)^n\rangle\;, \label{eq:approx}
\end{align}

i.e. the fluctuations of elliptic flow are mostly given by the mean response to initial state geometry fluctuations, while fluctuations in the response coefficient are strongly subdominant. We have further checked that if the mean is taken over centrality classes, such that the change of the response coefficient with centrality is taken into account, this statement remains true at finite shear viscosities. This fact allows to factorize the dependencies on initial state and flow response, which is what the derivation of the disentangling observables is based on.

\begin{figure}
    \centering
    \sidecaption
    \includegraphics[width=.5\linewidth]{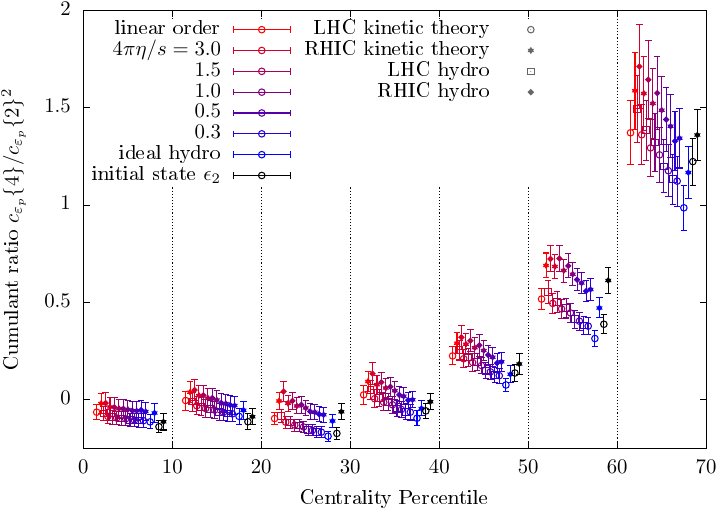}
    \caption{Ratio of the fourth order to the square of the second order elliptic flow cumulant as a function of centrality, as computed from hydrodynamic and kinetic theory based simulations of RHIC and LHC collisions (different symbols) at various values of the specific shear viscosity (different colors). Black points show the values for cumulants of the initial state ellipticity.}
    \label{fig:ratio}
    \vspace{-15pt}
\end{figure}

In order to construct an observable that is sensitive mostly to the initial state geometry, one has to get rid of the dependence on the flow response coefficient. This can be achieved by considering ratios of appropriate powers of flow cumulants \cite{Bhalerao:2011yg}, e.g.

\begin{align}
    \frac{c_{\varepsilon_p}\{4\}}{c_{\varepsilon_p}\{2\}^2}=\frac{\langle(\varepsilon_p)^4\rangle-2\langle(\varepsilon_p)^2\rangle^2}{\langle(\varepsilon_p)^2\rangle^2}\approx \frac{\langle(\epsilon_2)^4\rangle-2\langle(\epsilon_2)^2\rangle^2}{\langle(\epsilon_2)^2\rangle^2}=\frac{c_{\epsilon_2}\{4\}}{c_{\epsilon_2}\{2\}^2}\;.
\end{align}

As these observables give access only to ratios of initial state ellipticity cumulants, they cannot fully determine the initial state geometry. However, they do provide a way to directly measure its properties and constrain initial state models.

Fig.~\ref{fig:ratio} shows a verification of this fact using our simulation results. We plot the ratio $c_{\varepsilon_p}\{4\}/c_{\varepsilon_p}\{2\}^2$ as a function of centrality from different dynamical descriptions and for RHIC and LHC initial conditions as well as for different shear viscosities and compare to the ratio $c_{\epsilon_2}\{4\}/c_{\epsilon_2}\{2\}^2$ in black. Indeed all flow results are compatible with each other and agree with the initial state geometry ratio. The biggest difference in the results comes from the slightly different geometry in RHIC and LHC collision systems, but it is still small.

\begin{figure*}
    \centering
    \includegraphics[width=0.49\linewidth]{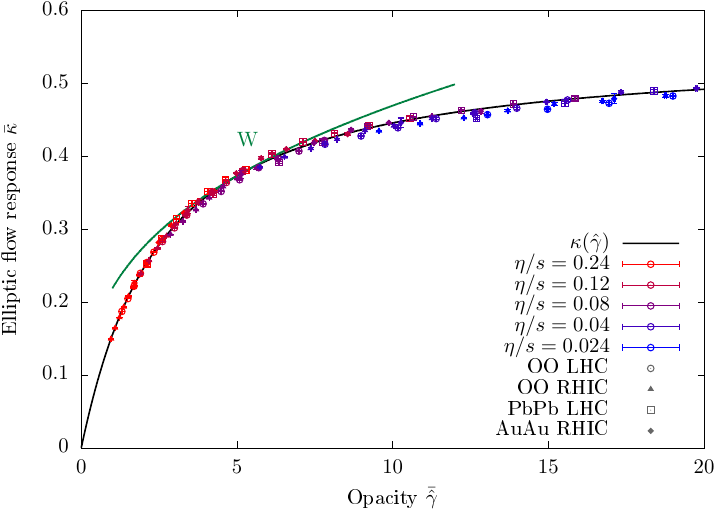}
    \includegraphics[width=0.49\linewidth]{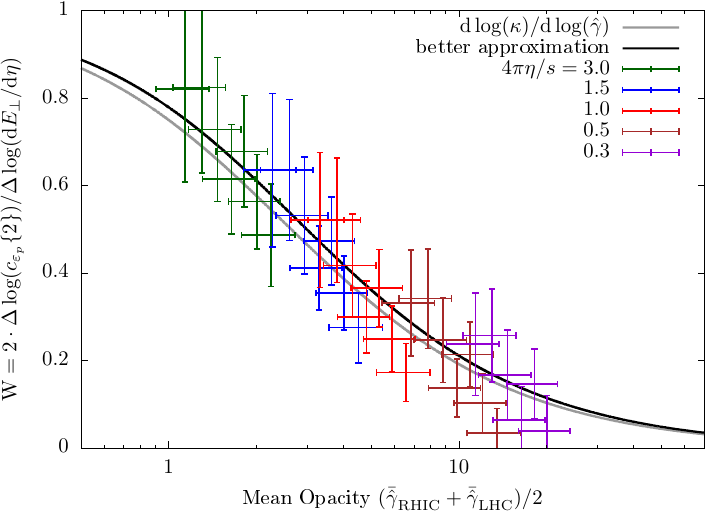}
    \caption{Left: Mean flow response coefficient $\bar{\kappa}$ as a function of the mean opacity $\bar{\hat{\gamma}}$. Points show results from the centrality classes of different collision systems (different symbols) from kinetic theory simulations. The black curve is the extracted fit for $\kappa(\hat{\gamma})$ and the dark green curve indicates how to obtain $\W$. Right: comparison of the extracted $\d\log\kappa/\d\log\hat{\gamma}$-curve (grey) and the improved theory expectation (black) to results for $\W$ from O+O simulations at different $\eta/s$ (different colors).}
    \label{fig:W_sim}
    \vspace{-20pt}
\end{figure*}

The first step in deriving a complementary observable that exclusively measures the flow response is to eliminate the dependence on the initial state geometry. In analogy to the observables that access the initial state geometry, one may construct ratios of flow cumulants to cancel the geometry factor. This requires to compare the centrality classes of two collision systems with roughly the same geometry but different scales of flow responses, for example two collision systems of the same nuclei but at different collisional energy, like O+O at RHIC and LHC. However, this will give access only to ratios of flow response coefficients between the two systems. In order to determine an absolute scale of flow responses, additional information has to be supplied. We therefore combine this ratio with the ratio of opacities in the two systems. If again the geometry (and shear viscosity) is similar, this is given by the ratio of transverse energies between the systems. We can thus measure the two ratios
\begin{align}
    \frac{\bar{\kappa}_{\rm RHIC}^{2k}}{\bar{\kappa}_{\rm LHC}^{2k}} \approx \frac{c_2^{\rm RHIC}\{2k\}}{c_2^{\rm LHC}\{2k\}} \quad \mathrm{and} \quad\frac{\bar{\hat{\gamma}}_{\rm RHIC}}{\bar{\hat{\gamma}}_{\rm LHC}} \approx \left(\frac{\langle{\d E_\perp}/{\d \eta}\rangle_{\rm RHIC}}{\langle{\d E_\perp}/{\d \eta}\rangle_{\rm LHC}}\right)^{1/4} \;.
\end{align}

These ratios can be turned into differences by using the logarithm. In order to connect to the flow response curve, we combine these two differences into the observable
\begin{align}
    \W=\frac{2}{k}\frac{\Delta\log(c_2\{2k\})}{\Delta\log(\d E_\perp/\d \eta)}\approx\frac{\d\log \kappa}{\d\log\hat{\gamma}}\;,
\end{align}
which is a finite difference approximation to the logarithmic derivative of the logarithmic flow response curve. More intuitively, this corresponds to the local power law of the $\kappa(\hat{\gamma})$-curve, as indicated by the green curve in the left plot of Fig.~\ref{fig:W_sim}. Thus, using knowledge of the flow response curves, a measurement of $\W$ allows to assess the typical opacity and therefore the degree of hydrodynamization as a mean of the two systems.

We now aim to verify that $\W$ works as intended on the basis of our simulation results. To this end, we extract the elliptic flow response curve $\kappa(\hat{\gamma})$ as a fit to centrality class averages of the event-by-event values of $\kappa=\varepsilon_p/\epsilon_2$ and the opacity $\hat{\gamma}$, as shown in the left plot of Fig.~\ref{fig:W_sim}. Here, we also included results from simulations of Pb+Pb collisions at LHC and Au+Au collisions at RHIC. The resulting $\d\log\kappa/\d\log\hat{\gamma}$-curve is shown in grey in the right plot of Fig. \ref{fig:W_sim}. The black curve slightly improves on it by taking into account also the difference between initial state and final state transverse energies. As expected from the discussion in the introduction, the curve monotonically transitions from a value of 1 in the low opacity regime to the ideal hydrodynamic limit of 0. Thus, it provides a one-to-one correspondence between the hydrodynamization observable $\W$ and the opacity $\hat{\gamma}$.

We further compute $\W$ from our simulation data for $c_{\varepsilon_p}\{2\}$ and $\d E_\perp/\d \eta$ and compare to $\d\log\kappa/\d\log\hat{\gamma}$ in the right plot of Fig. \ref{fig:W_sim}. Again, varying the shear viscosity allows to scan a large range in opacity using the same geometry. All extracted values of $\W$ agree well with the theory expectation. Thus, on the level of our simulation results, $\W$ works as intended.

\begin{wrapfigure}{r}{.45\textwidth}
\vspace{-5pt}
    \centering
    \includegraphics[width=\linewidth]{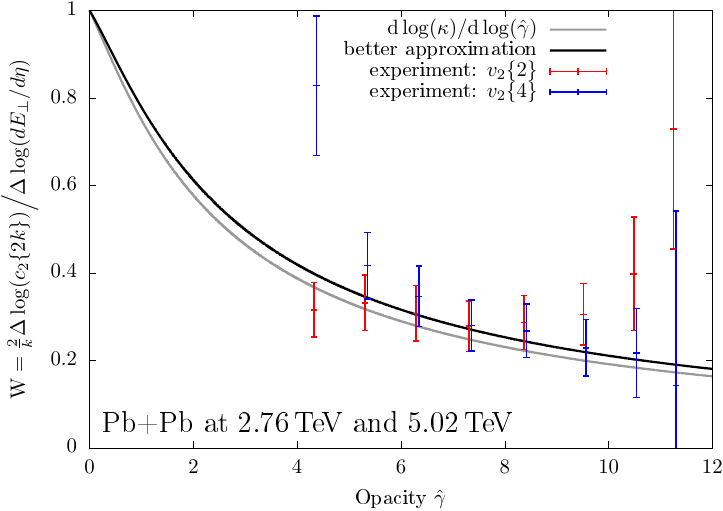}
    \caption{$\W$ as computed from experimental data for $v_2\{2\}$ (red) and $v_2\{4\}$ (blue) in Pb+Pb collisions at LHC, 
    compared to the two theory expectation curves (grey and black).}
    \label{fig:W_exp}
    \vspace{-15pt}
\end{wrapfigure}

Finally, as a first test of how the observable $\W$ fares in practice, we computed it from measurements of Pb+Pb collisions at the LHC, combining results from runs at $2.76\;$TeV and $5.02\;$TeV. To this end, we reconstructed of the second and fourth order flow cumulants from $v_2\{2\}$ and $v_2\{4\}$ for several centrality classes. The results for $\W$ are compared to the $\d\log\kappa/\d\log\hat{\gamma}$-expectation in Fig.~\ref{fig:W_exp}. For the positioning on the $\hat{\gamma}$-axis, the mean opacities of the two Pb+Pb systems were estimated using \trento~initial conditions and a value of the shear viscosity $\eta/s=0.12$, at which simulation results for the transverse energy agree with experimental measurements. Both extractions of $\W$ are compatible with the theory expectation. As a function of opacity, the results from $v_2\{2\}$ are flatter than expected. This might be due to non-flow effects, as the results from $v_2\{4\}$ seem to also follow the opacity dependence of $\d\log\kappa/\d\log\hat{\gamma}$. We conclude that indeed $\W$ has the potential to allow for experimental measurements of the degree of hydrodynamization in hadronic collision systems.

\textbf{Acknowledgments:}
This work is supported by the Deutsche Forschungsgemeinschaft (DFG, German Research Foundation)
through the CRC-TR 211 ’Strong-interaction matter under extreme conditions’– project
number 315477589 – TRR 211.
V.E.A. gratefully acknowledges support by the European Union - NextGenerationEU through grant No. 760079/23.05.2023, funded by the Romanian ministry of research, innovation and digitization through Romania’s National Recovery and Resilience Plan, call no.~PNRR-III-C9-2022-I8.
C.W. was supported by the program Excellence Initiative–Research University of the University of Wrocław of the Ministry of Education and Science.
C.W. has also received funding from the European Research Council (ERC) under the European Union’s Horizon 2020 research and innovation programme (grant number: 101089093 / project acronym: High-TheQ). Views
and opinions expressed are however those of the authors
only and do not necessarily reflect those of the European
Union or the European Research Council. Neither the
European Union nor the granting authority can be held
responsible for them.
\vspace{-15pt}

%
 \bibliography{your_bib_file} 
%
%
%
%

\end{document}